\documentclass[aps,prl,reprint,superscriptaddress,longbibliography]{revtex4-1}
\usepackage[latin1]{inputenc}
\usepackage[T1]{fontenc}
\usepackage{amsmath}
\usepackage{amsfonts}
\usepackage{amssymb}
\usepackage{enumerate}
\usepackage{amscd}
\usepackage{amsthm}
\usepackage[english]{babel}
\usepackage[pdftex]{graphicx}
\usepackage[pdftex]{thumbpdf}
\usepackage[bookmarks=true,pdfpagemode=UseOutlines]{hyperref}
\usepackage{bookmark}
\usepackage{fancyhdr}
\usepackage{float}
\usepackage{array}
\usepackage{dcolumn}

\begin{document}

\title{Nature of the spin liquid ground state in a breathing kagome compound studied by NMR and series expansion}
\date{\today}
\author{J.-C. Orain}
\affiliation{Laboratoire de Physique des Solides, CNRS, Univ. Paris-Sud, Universit\'e Paris-Saclay, 91405 Orsay Cedex, France}
\affiliation{Laboratory for Muon Spin Spectroscopy, Paul Scherrer Insititut, 5232 Villigen PSI, Switzerland}
\author{B. Bernu}
\affiliation{LPTMC, UMR 7600 of CNRS, UPMC, Paris-Sorbonne, F-75252 Paris Cedex 05, France}
\author{P. Mendels}
\affiliation{Laboratoire de Physique des Solides, CNRS, Univ. Paris-Sud, Universit\'e Paris-Saclay, 91405 Orsay Cedex, France}
\author{L. Clark}
\affiliation{School of Chemistry and EaStChem, University of St Andrews, St Andrews, Fife, KY16 9ST, UK}
\affiliation{Departments of Chemistry and Physics, School of Physical Sciences, University of Liverpool, Liverpool, L69 7ZD (UK)}
\author{F.~H. Aidoudi}
\affiliation{School of Chemistry and EaStChem, University of St Andrews, St Andrews, Fife, KY16 9ST, UK}
\author{P. Lightfoot}
\affiliation{School of Chemistry and EaStChem, University of St Andrews, St Andrews, Fife, KY16 9ST, UK}
\author{R.~E. Morris}
\affiliation{School of Chemistry and EaStChem, University of St Andrews, St Andrews, Fife, KY16 9ST, UK}
\author{F. Bert}
\email{fabrice.bert@u-psud.fr}
\affiliation{Laboratoire de Physique des Solides, CNRS, Univ. Paris-Sud, Universit\'e Paris-Saclay, 91405 Orsay Cedex, France}

\begin{abstract}
In the vanadium oxyfluoride compound (NH$_4$)$_2$[C$_7$H$_{14}$N][V$_7$O$_6$F$_{18}$] (DQVOF), the V$^{4+}$ (3d$^1$, $S=1/2$) ions realize a unique, highly frustrated breathing kagome lattice composed of alternately-sized, corner-sharing equilateral triangles. Here we present an  $^{17}$O NMR study of DQVOF, which isolates the local susceptibility of the breathing kagome network. By a fit to series expansion we extract the ratio of the interactions within the breathing kagome plane, $J_\triangledown / J_\vartriangle = 0.55(4)$, and the mean antiferromagnetic interaction $\bar{J}=60(7)$~K. Spin lattice, $T_1$, measurements reveal an essentially gapless excitation spectrum with a maximum gap $\Delta / \bar{J}=0.007(7)$. Our study provides new impetus for further theoretical investigations in order to establish whether the gapless spin liquid behavior displayed by DQVOF  is intrinsic to its breathing kagome lattice or whether it is due to perturbations to this model, such as a residual coupling of the V$^{4+}$ ions in the breathing kagome planes to the interlayer V$^{3+}$ ($S=1$) spins.
\end{abstract}

\maketitle

Frustrated magnetism stands at the forefront of theoretical and experimental quantum matter research and has emerged as a vast playground in which to realize novel exotic states~\cite{Balents2010}. One emblematic example is the antiferromagnetic Heisenberg model for $S=1/2$ spins on the kagome lattice (KAFM) made of corner-sharing triangles. Despite its seemingly simple nature, the $S=1/2$ KAFM continues to provide a rich source of novel physics~\cite{Ran2007,Messio2012} and  challenges the most advanced analytical and numerical methods~\cite{Yan2011,Depenbrock2012,Liao2016}. The complexity of this problem stems from the many ground state candidates - mostly quantum spin liquid phases for which there is no associated symmetry breaking of the Hamiltonian - which compete within a narrow range of energy. Even the existence of a gap in the excitation spectrum is still hotly debated~\cite{Nakano2011,Iqbal2013,Iqbal2014,Liao2016}. In parallel, advances in materials science have led to the identification of several spin liquid candidates among which herbertsmithite, ZnCu$_3$(OH)$_6$Cl$_2$, realizes the closest experimental approximation to the KAFM~\cite{Shores2005,Mendels2011,Norman2016}. However, the exact nature of the ground state of herbertsmithite remains unclear. First, the presence of a sizeable amount of Cu$^{2+}$ ions sitting out of the kagome layers pollutes the low temperature magnetic properties and renders it challenging to definitively conclude whether or not there is a spin gap of few kelvins in the excitation spectrum~\cite{Olariu2008,Fu2015,Han2012a}. Second,  the absence of a firm theoretical prediction for the ideal  KAFM model does not allow for a deep quantitative analysis nor for an evaluation of how perturbations from the ideal  KAFM Hamiltonian that inevitably exist in real materials, such as defects, anisotropic interactions or further nearest neighbor interactions, influence the nature of the ground state.

A modified version of the kagome lattice, the \emph{trimerized}, or \emph{breathing} kagome lattice~\cite{Schaffer17} named after the breathing pyrochlores~\cite{Okamoto2013}, may offer an alternative route for the study of quantum spin liquids. Importantly, it also consists of corner-sharing equilateral triangles and hence it retains the full degeneracy of the kagome lattice at the classical level. However the alternation of the interactions within the triangles pointing up, $J_\vartriangle$, and those pointing down, $J_\triangledown$, (see Fig 1, inset) extends the problem of the isotropic KAFM ($J_\vartriangle=J_\triangledown$) to a larger class of highly frustrated systems. The trimerization may then select more clearly one type of spin liquid for the ground state~\cite{Schaffer17}. This would constitute a clear theoretical standpoint to investigate the spin liquid physics in the recently discovered vanadium oxyfluoride compound (NH$_4$)$_2$[C$_7$H$_{14}$N][V$_7$O$_6$F$_{18}$] (DQVOF)~\cite{Aidoudi2011} which features such a breathing kagome lattice and which is the focus of the present study.

\begin{figure}
\centering
\includegraphics[scale=0.2]{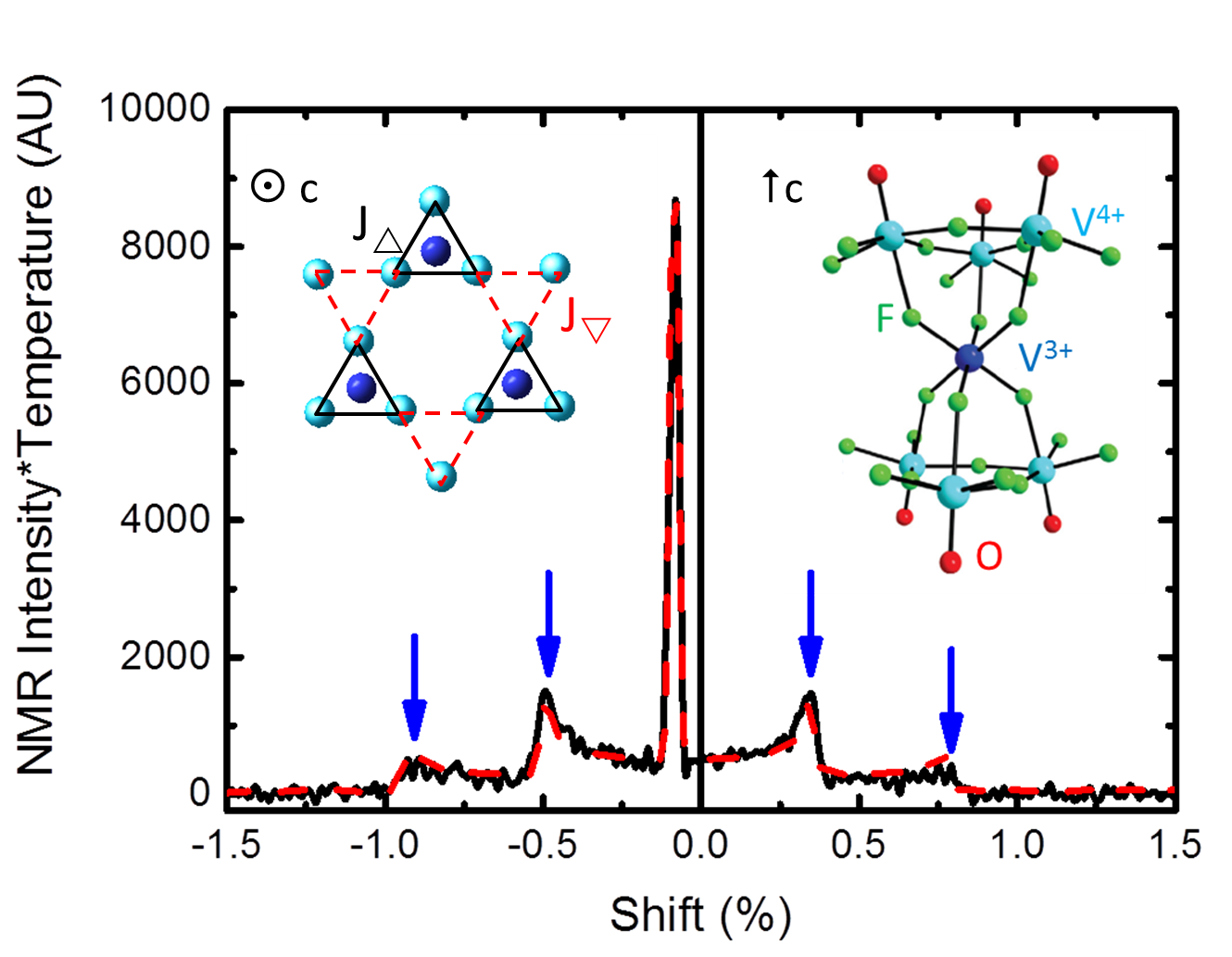}
\caption{\label{HTOxy} $^{17}$O NMR spectrum of DQVOF at 300~K and $H_0=7.5$~T (black line) and the simulation of the powder spectrum (red dashed line). The blue arrows point to the quadrupolar singularities. The shift reference is $\nu_0=$~$^{17}\gamma /2 \pi \times H_0=43.288$~MHz ($^{17} \gamma / 2 \pi=5.772$~MHz/T). Left inset: top view of the magnetic lattice with the V$^{4+}$ (light blue) forming the breathing kagome lattice and the interlayer V$^{3+}$ (dark blue). Right inset: local environments of the vanadium ions.}
\end{figure}

The structure of DQVOF consists of magnetic trilayer well separated by non-magnetic quinuclidinium [C$_7$H$_{14}$N]$^+$ cations. Each trilayer contains two V$^{4+}$ (3d$^1$, $S=1/2$) kagome layers separated by a triangular interlayer of V$^{3+}$ (3d$^2$, $S=1$) ions (see insets in Fig.~\ref{HTOxy}). The V$^{3+}$ ions are located above or below every other triangle of V$^{4+}$ ions, which yields the two set of alternating equilateral triangles within the $S=1/2$ breathing kagome lattice. All the vanadium ions are coupled by superexchange via fluoride ions which leads to antiferromagnetic Curie-Weiss behavior at high temperature with a Weiss temperature $\theta=58(4)$~K. Due to the compression of the octahedral environment of the V$^{4+}$, the $d_{xy}$ orbital of the V$^{4+}$ is singly occupied~\footnote{The VOF$_5$ octahedron is strongly distorted with a distance $d_{V^{4+}-O}=1.58$~\AA{} much shorter than the distances $d_{V^{4+}-F}$ all in the range $1.95-2.15$~\AA{} which lifts the $t_{2g}$ orbitals degeneracy and favors the $d_{xy}$ one~\cite{Aidoudi2011}.}. This lies close to the kagome plane and, therefore, little coupling is expected in the perpendicular direction to the V$^{3+}$ ions. Indeed, previous magnetization and heat capacity measurements~\cite{Clark2013} revealed the paramagnetic-like response of the V$^{3+}$ ions, which have a weak residual effective coupling $\sim~1$~K . As such, the $S=1/2$ kagome layers appear to be magnetically isolated from one another. Muon spin relaxation experiments down to 20~mK revealed no sign of any on-site frozen moments~\cite{Clark2013,Orain2014} as expected for a spin liquid ground state.

In this Letter, we report the $^{17}$O nuclear magnetic resonance (NMR) study of DQVOF which isolates the local magnetic properties of the V$^{4+}$ breathing kagome layers. The comparison between the local susceptibility and our theoretical calculations based on series expansion allows us to determine the ratio of the interactions $J_\triangledown / J_\vartriangle \sim 0.55$ while the dynamical spin lattice relaxation ($T_1$) measurements reveal a gapless excitation spectrum.

A polycrystalline sample (84~mg) was synthesized as described in \cite{Aidoudi2011} with the addition of 50~$\mu$l of 90~\% $^{17}$O enriched water in the ionic solution~\cite{Griffin2012}. The NMR spectra were recorded from 10 - 300~K in a fixed field of 7.5~T by sweeping the frequency, while the broader spectra below 10~K were measured with a fixed irradiating frequency by sweeping the field from 6.5~T to 7.15~T and integrating the spin echo.

The $^{17}$O NMR spectrum at 300~K (Fig \ref{HTOxy}) is a typical powder spectrum for an $I=5/2$ nuclear spin in an almost axially symmetric environment. It consists of one sharp central line, \emph{i.e.} one oxygen site as expected from the crystal structure, in between two pairs of well defined quadrupolar singularities. The observation of one single central line together with the small distribution of the quadrupolar parameters~\footnote{From a powder spectrum simulation, we extract an average quadrupolar frequency $\nu_Q=370$~kHz with a small distribution $\Delta \nu_Q=10$~kHz and a maximal anisotropy of the quadrupolar tensor $\eta <0.05$. The shift tensor $\bar{\mathbf{K}}$ is almost isotropic with $K_{x,y}/K_z=0.87$, negligible in view of the low $T$ broadening.} demonstrates the absence of any sizable disorder and, in particular, of any magnetic dilution of the $S=1/2$ kagome layers.

\begin{figure}
\centering
\includegraphics[width=\columnwidth]{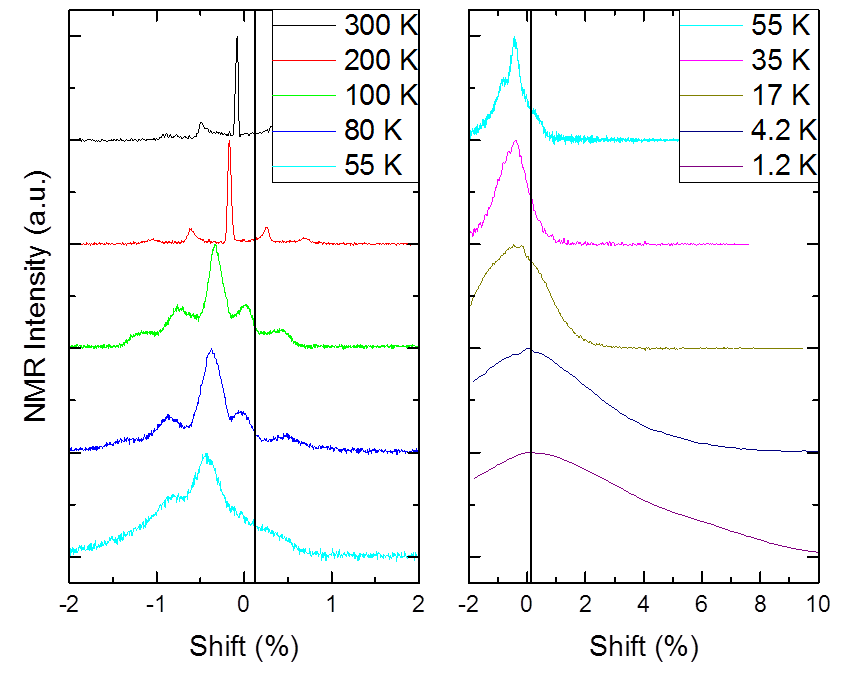}
\caption{\label{Cascade} $^{17}$O NMR spectra over a selected range of temperatures. The spectra are normalized to their maximum value. The black vertical lines show the reference for the spin shift. }
\end{figure}

\begin{figure}
\centering
\includegraphics[scale=0.35]{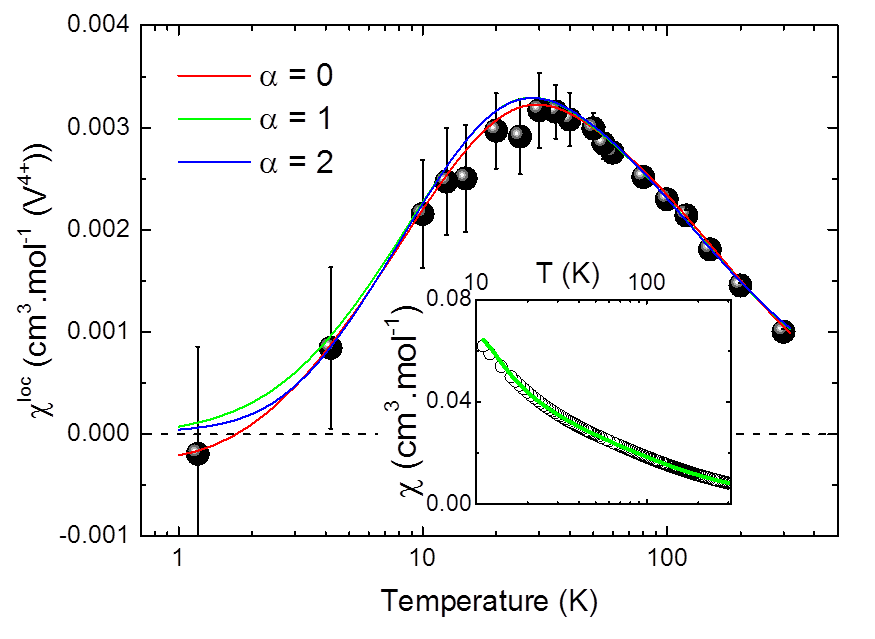}
\caption{\label{Fit} Local NMR susceptibility of the kagome layers $\chi^{loc}$. The lines are fits to $A\chi^{th}_{kago}+C$ (see text). Inset: SQUID macroscopic susceptibility (circles) and fit to a combination of $\chi^{th}_{kago}$ ($\alpha=1$) and a paramagnetic contribution from the V$^{3+}$ (line)~\cite{SI}.}
\end{figure}

Upon cooling the sample (Fig \ref{Cascade}), the NMR spectrum first shifts negatively away from the reference line down to approximately 30~K. At lower temperature, it shifts back towards the reference and broadens drastically. The NMR shift is related to the local spin susceptibility $\chi^{loc}$ through $K=\mathcal{A}\chi^{loc}(T)+K_0$ where $K_0$ is the $T$-independent shift including the orbital shift and $\mathcal{A}$ is the isotropic hyperfine coupling. Given that the oxide ions are directly bonded to the V$^{4+}$ ions within the DQVOF structure at a distance of 1.58~\AA~and are far away from the other magnetic centers, there is little doubt that the $^{17}$O nuclei are hyperfine coupled to those V$^{4+}$ and thus that the lineshift predominatly probes the local susceptibility of the $S=1/2$ kagome layer. Note that, being a traceless tensor, the dipolar field arising from the polarization of the V$^{3+}$ does not contribute to the lineshift in a powder sample but to the linewidth, yielding in particular the low $T$ broadening when the magnetization of the quasi-free V$^{3+}$ increases drastically~\cite{SI}. A plot of the high temperature shift ($T>80$~K) versus a Curie-Weiss model for the V$^{4+}$ kagome susceptibility~\cite{SI} gives $\mathcal{A} = -11.0(5)$~kOe.$\mu_B^{-1}$ and $K_0=1200(200)$~ppm. The local spin susceptibility $\chi^{loc}$ of the kagome layers depicted in Fig.~\ref{Fit} reveals a broad maximum at 30~K$\approx \theta /2$, in striking contrast to the monotonic variation of the macroscopic susceptibility (see inset, Fig.~\ref{Fit}) which, unlike the local susceptibility measured by $^{17}$O NMR, encompasses the contribution of the quasi-free V$^{3+}$ spins. The decrease of $\chi^{loc}$ at low temperatures provides evidence for the enhancement of the short-range antiferromagnetic spin correlations without long-range magnetic order which is in-keeping with the spin liquid description of the ground state~\cite{Clark2013}. The overall behavior of the local susceptibility of DQVOF is in fact very similar to that reported for herbertsmithite~\cite{Fu2015}.

We now turn to compare the experimentally determined $\chi^{loc}$ with the susceptibility $\chi^{th}_{kago}$ of the $S=1/2$ Heisenberg model on the breathing kagome lattice, calculated over the whole temperature range using the series expansion method described in Ref.~\cite{Bernu2015,SI}. In brief, this method relies on the extrapolation of the entropy $s(e,H)$ for an applied magnetic field, $H$, over the whole range of energies, $e$,  starting from the high temperature series expansion of $s(T)$ and $e(T)$. The free energy $f=e-Ts$ is then computed at all $T$ and yields the susceptibility
\begin{equation}
\nonumber
\chi^{th}_{kago}(T)=\frac{1}{H}\left( \frac{\partial f}{\partial H} \right )_T.
\end{equation}
The extrapolation of $s(e,H)$ at low energy depends on the \textit{ad hoc} choice of the type of behavior of the specific heat $C_V(T)$ at $T\rightarrow0$. We have tried three cases as shown on Fig.~\ref{Fit}: $i)$ gapped, hereafter noted $\alpha=0$ (red line), $ii)$ power laws : $C_V\propto T^\alpha$ with $\alpha=1$ (green line) and $2$ (blue line). The theoretical curves were fitted to the experimental data in order to minimize the fit error $E$ defined as
\begin{eqnarray}
\nonumber
E=\frac{1}{\epsilon n_T}\sum_{i=1}^{n_T}T_i\left(A\chi^{th}_{kago}(T_i)+C-\chi^{loc}(T_i) \right)^2
\end{eqnarray}
where $n_T$ is the number of data points, $T_i$ the data temperatures, $\epsilon=0.0012$ an estimate of the experimental error on $T (\chi^{loc})^2$, $A\simeq 1$  and $C$ are corrections to the experimentally determined hyperfine coupling and $T$-independent shift, respectively. In view of the experimental data, we further imposed that $\chi^{th}_{kago}(T=0)=0$.

The fitting parameters obtained in this way are summarized in Table~\ref{ValueFit}. We found fits of similarly high quality for the three possible values of $\alpha$ and it is thus impossible to conclude on the existence of a spin gap on the sole basis of the susceptibility data.  In the gapped case ($\alpha=0$), the gap value refines to $\Delta=3(1)$~K~$=0.05(2) \bar{J}$, in the range of the currently debated DMRG results for the isotropic kagome lattice~\cite{Depenbrock2012,Nishimoto13,He16}. Former analysis of the low temperature heat capacity of DQVOF, after subtraction of field dependent contributions, rather supports the linear $\alpha=1$ case~\cite{Clark2013}. Nonetheless, the various fits give a consistent determination of the breathing ratio $J_\triangledown/J_\vartriangle \simeq 0.55(4)$ providing firm ground for further numerical studies based on this ratio. Note  that enforcing an isotropic model $J_\triangledown=J_\vartriangle$ yields unphysical parameters~\cite{SI}. Moreover, our attempts to fit the total SQUID susceptibility in the range 20 - 300~K by adding a simple Curie-Weiss contribution for the interlayer V$^{3+}$ (Fig.~\ref{Fit}, inset and \cite{SI}) gave a reasonably consistent value for the ratio of the interactions $J_\triangledown/J_\vartriangle \simeq 0.45(12)$.

\begin{table*}
\caption{\label{ValueFit}Parameters of the series expansion fits of the NMR local susceptibility (see text).}
\begin{ruledtabular}
\begin{tabular}{c|cccc|c|cc}
$\alpha$ & $J_\vartriangle\text{ (K) }$ & $J_\triangledown\text{ (K) }$ & $J_\triangledown /J_\vartriangle$ & $\bar{J}=(J_\triangledown +J_\vartriangle)/2\text{ (K) }$ & $\Delta$~(K) & $A$ & $C\text{ (cm}^{3}.\text{mol}^{-1}\text{ (V}^{4+}\text{)) }$ \\ \hline
0 & 84(1) & 49(1) & 0.58 & 67(2) & 3(1) & 1.31 & -2.6.10$^{-4}$ \\
1 & 73(1) & 37(1) & 0.51 & 55(2) &  --- & 1.06 & -2.11.10$^{-5}$ \\
2 & 68(1) & 37(1) & 0.54 & 53(2) & --- & 1.02 & 1.92.10$^{-5}$
\end{tabular}
\end{ruledtabular}
\end{table*}


\begin{figure}
\centering
\includegraphics[scale=0.3]{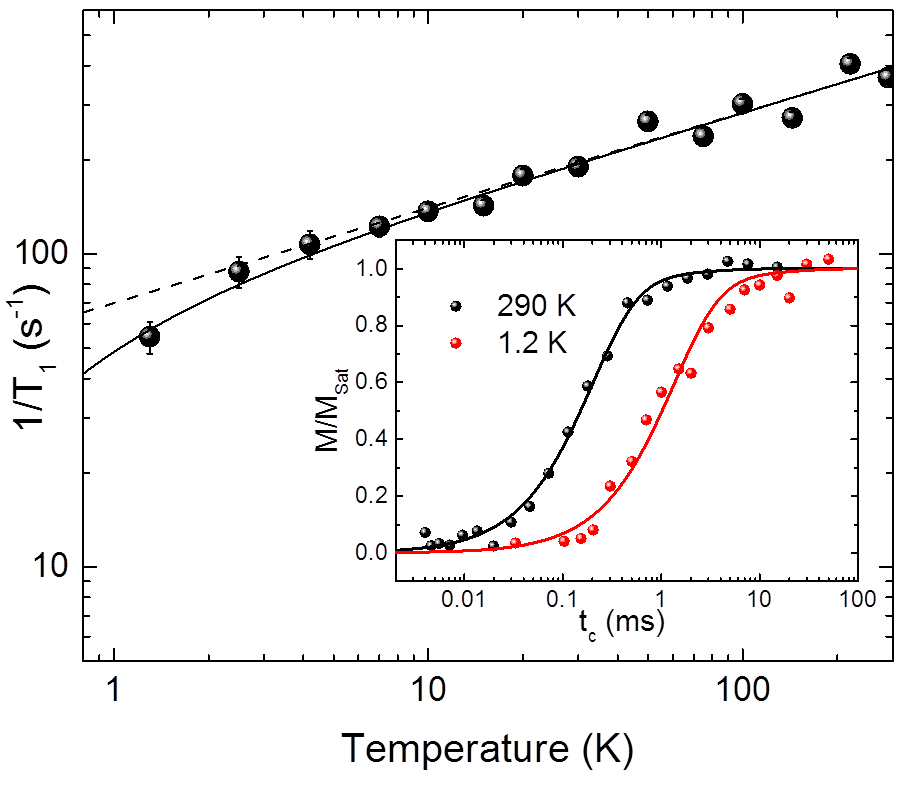}
\caption{\label{T1} Thermal evolution of $1/T_1$. The solid line is a fit as explained in the text. The dashed line shows the $T^{0.3}$ variation. Inset: recovery curves at 290~K and 1.2~K. }
\end{figure}

In order to investigate the possible presence of a spin gap in DQVOF more precisely, we have performed NMR spin lattice ($T_1$) relaxation measurements which probe the spin excitation spectrum through $1/T_1 \propto T \chi''(\omega_n,T)/\omega_n$ where $\omega_n / 2 \pi=~^{17}\gamma H_0 / 2 \pi$ is the NMR frequency. The $1/T_1$ relaxation rates were measured at the top of the central line using a saturating pulse sequence $\pi/2$-$t_c$-$\pi/2$-$\tau$-$\pi$. The recovery curves (Fig.~\ref{T1}, inset) could be fitted over the whole $T$ range using the expression
\begin{equation}
\nonumber
\frac{M(t_c)}{M_{sat}}=1-\frac{1}{35}e^{-t_c/T_1}+\frac{8}{45}e^{-6t_c/T_1}+\frac{50}{63}e^{-15t_c/T_1}
\end{equation}
corresponding to the irradiation of the central line of an $I=5/2$ nucleus~\cite{Suter1998}. Hence a single $T_1$ is determined unambiguously at all $T$ and the absence of distribution of $T_1$ down to the base temperature points to a homogeneous ground state. Note that the V$^{3+}$ ions are fully polarized at low $T$ in the 7~T applied field~\cite{Clark2013} and therefore do not pollute the measurements of the kagome spin relaxation. The thermal evolution of $T_1$ is remarkably featureless, showing a weak power law increase $T^{\sim 0.3}$ over two decades in temperatures encompassing the energy scale of the interactions $\bar{J}$. At 1.2~K the $1/T_1$ value falls short of this behavior and may suggest a change towards slower dynamics. To capture this possible change of regime we have fitted the $1/T_1$ data over the whole $T$ range to the phenomenological function $1/T_1 \propto T^\beta \exp (-\Delta/T )$, which accounts for a crossover from a pure power law behavior at high temperature to a gapped behavior at low $T$. We obtained  $\Delta=0.4(4)$~K$\approx 0.007(7)~\bar{J}$ and $\beta=0.30(3)$.
The  vanishingly small value of $\Delta$, well below the estimates of the gap 0.05~J-0.13~J for gapped spin liquids in the isotropic model, points to an essentially gapless excitation spectrum for the breathing kagome lattice within DQVOF, in line with the linear $T$-dependence of the heat capacity suggesting gapless fermionic spinon excitations~\cite{Clark2013}.

We now consider the effect of the breathing of the kagome lattice on the spin gap issue. For nearly isolated spin clusters, one expects a large spin gap, $\sim J_\vartriangle$, from the discrete energy spectrum as observed for instance in LiInCr$_4$O$_8$ that features well decoupled tetrahedral clusters~\cite{Okamoto2013}. Increasing the intercluster interaction lifts the degeneracy of the individual ground states. A small gap, $\ll J_\vartriangle$, may then potentially open as suggested in early studies in the strong coupling regime ($ J_\triangledown \ll J_\vartriangle $)~\cite{Mila1998,Mambrini2000,Zhitomirsky2005a}. A recent variational Monte Carlo study suggested a gapped state also for $0.5<J_\triangledown /J_\vartriangle <0.8$~\cite{Schaffer17}. The evidence for a gapless ($\Delta<0.007(7) \bar{J}$) spin liquid in DQVOF clearly calls for further theoretical investigation into the breathing KAFM including the calculation of the spin gap value, if any. Understanding how the spin liquid physics evolves with breathing and eventually connects to the isotropic kagome case is \emph{per se} an interesting theoretical issue. Furthermore, in DQVOF, the presence of V$^{3+}$ interlayer ions is likely the dominant perturbation to be considered beyond the  breathing kagome model. This situation bears some interesting similarities with the widely-debated Cu$^{2+}$ interlayer defects in herbertsmithite. The relevant energy scale of this perturbation is the coupling between the interlayer spins and the spins in the kagome planes. In both cases this interaction is likely to be weak but difficult to estimate. Earlier studies suggested an effective $\sim$1~K interaction~\cite{Clark2013,Orain2014} which likely involves the kagome layer spins since no obvious exchange path exists between the V$^{3+}$ ions. We found, in particular, that their contribution to the low temperature susceptibility cannot be described satisfactorily by a simple Curie-Weiss model~\cite{SI}. The effect of such 'dangling' spins has not been widely investigated so far, but may yield rich physics encompassing the Kondo effect~\cite{Kolezhuk2006,Gomilsek2016} or subtle correlation effects through the kagome layer~\cite{Han2016}.

In conclusion, using NMR techniques we have determined the local $S=1/2$ V$^{4+}$ susceptibility of DQVOF, which can be accounted for by series expansion for the breathing kagome model. The determination of the ratio of the interactions within the breathing kagome lattice, $J_\triangledown /J_\vartriangle = 0.55(4)$, will allow for further theoretical investigations into the effects of the breathing of the $S=1/2$ KAFM lattice that are specific to DQVOF. It is hoped that this in turn will shed light on our experimental observation of a gapless spin liquid ground state. The breathing of the kagome lattice appears to be an interesting tuning parameter that may be used to control the spin liquid physics and one that warrants further exploration. Investigating new anisotropic kagome compounds with different layered structures is also a promising experimental route to furthering our understanding of the rich physics of the $S=1/2$ KAFM~\cite{Clark2015}.

\begin{acknowledgments}
This work was supported by the French Agence Nationale de la Recherche under Grant No. ANR-12-BS04-0021 'SPINLIQ'. J. C. Orain acknowledges support from the Region Ile de France through the DIM OxyMore. Work at the University of St Andrews was supported by the Leverhulme Trust RPG-2013-343.
\end{acknowledgments}


\begin{thebibliography}{35}%
\makeatletter
\providecommand \@ifxundefined [1]{%
 \@ifx{#1\undefined}
}%
\providecommand \@ifnum [1]{%
 \ifnum #1\expandafter \@firstoftwo
 \else \expandafter \@secondoftwo
 \fi
}%
\providecommand \@ifx [1]{%
 \ifx #1\expandafter \@firstoftwo
 \else \expandafter \@secondoftwo
 \fi
}%
\providecommand \natexlab [1]{#1}%
\providecommand \enquote  [1]{``#1''}%
\providecommand \bibnamefont  [1]{#1}%
\providecommand \bibfnamefont [1]{#1}%
\providecommand \citenamefont [1]{#1}%
\providecommand \href@noop [0]{\@secondoftwo}%
\providecommand \href [0]{\begingroup \@sanitize@url \@href}%
\providecommand \@href[1]{\@@startlink{#1}\@@href}%
\providecommand \@@href[1]{\endgroup#1\@@endlink}%
\providecommand \@sanitize@url [0]{\catcode `\\12\catcode `\$12\catcode
  `\&12\catcode `\#12\catcode `\^12\catcode `\_12\catcode `\%12\relax}%
\providecommand \@@startlink[1]{}%
\providecommand \@@endlink[0]{}%
\providecommand \url  [0]{\begingroup\@sanitize@url \@url }%
\providecommand \@url [1]{\endgroup\@href {#1}{\urlprefix }}%
\providecommand \urlprefix  [0]{URL }%
\providecommand \Eprint [0]{\href }%
\providecommand \doibase [0]{http://dx.doi.org/}%
\providecommand \selectlanguage [0]{\@gobble}%
\providecommand \bibinfo  [0]{\@secondoftwo}%
\providecommand \bibfield  [0]{\@secondoftwo}%
\providecommand \translation [1]{[#1]}%
\providecommand \BibitemOpen [0]{}%
\providecommand \bibitemStop [0]{}%
\providecommand \bibitemNoStop [0]{.\EOS\space}%
\providecommand \EOS [0]{\spacefactor3000\relax}%
\providecommand \BibitemShut  [1]{\csname bibitem#1\endcsname}%
\let\auto@bib@innerbib\@empty
\bibitem [{\citenamefont {Balents}(2010)}]{Balents2010}%
  \BibitemOpen
  \bibfield  {author} {\bibinfo {author} {\bibfnamefont {L.}~\bibnamefont
  {Balents}},\ }\bibfield  {title} {\enquote {\bibinfo {title} {{Spin liquids
  in frustrated magnets.}}}\ }\href@noop {} {\bibfield  {journal} {\bibinfo
  {journal} {Nature}\ }\textbf {\bibinfo {volume} {464}},\ \bibinfo {pages}
  {199--208} (\bibinfo {year} {2010})}\BibitemShut {NoStop}%
\bibitem [{\citenamefont {Ran}\ \emph {et~al.}(2007)\citenamefont {Ran},
  \citenamefont {Hermele}, \citenamefont {Lee},\ and\ \citenamefont
  {Wen}}]{Ran2007}%
  \BibitemOpen
  \bibfield  {author} {\bibinfo {author} {\bibfnamefont {Y.}~\bibnamefont
  {Ran}}, \bibinfo {author} {\bibfnamefont {M.}~\bibnamefont {Hermele}},
  \bibinfo {author} {\bibfnamefont {P.~A.}\ \bibnamefont {Lee}}, \ and\
  \bibinfo {author} {\bibfnamefont {X.~G.}\ \bibnamefont {Wen}},\ }\bibfield
  {title} {\enquote {\bibinfo {title} {{Projected-Wave-Function Study of the
  Spin-1/2 Heisenberg Model on the Kagome Lattice}},}\ }\href@noop {}
  {\bibfield  {journal} {\bibinfo  {journal} {Phys. Rev. Lett.}\ }\textbf
  {\bibinfo {volume} {98}},\ \bibinfo {pages} {117205} (\bibinfo {year}
  {2007})}\BibitemShut {NoStop}%
\bibitem [{\citenamefont {Messio}\ \emph {et~al.}(2012)\citenamefont {Messio},
  \citenamefont {Bernu},\ and\ \citenamefont {Lhuillier}}]{Messio2012}%
  \BibitemOpen
  \bibfield  {author} {\bibinfo {author} {\bibfnamefont {L.}~\bibnamefont
  {Messio}}, \bibinfo {author} {\bibfnamefont {B.}~\bibnamefont {Bernu}}, \
  and\ \bibinfo {author} {\bibfnamefont {C.}~\bibnamefont {Lhuillier}},\
  }\bibfield  {title} {\enquote {\bibinfo {title} {Kagome antiferromagnet: A
  chiral topological spin liquid?}}\ }\href {\doibase
  10.1103/PhysRevLett.108.207204} {\bibfield  {journal} {\bibinfo  {journal}
  {Phys. Rev. Lett.}\ }\textbf {\bibinfo {volume} {108}},\ \bibinfo {pages}
  {207204} (\bibinfo {year} {2012})}\BibitemShut {NoStop}%
\bibitem [{\citenamefont {Yan}\ \emph {et~al.}(2011)\citenamefont {Yan},
  \citenamefont {Huse},\ and\ \citenamefont {White}}]{Yan2011}%
  \BibitemOpen
  \bibfield  {author} {\bibinfo {author} {\bibfnamefont {S.}~\bibnamefont
  {Yan}}, \bibinfo {author} {\bibfnamefont {D.~A.}\ \bibnamefont {Huse}}, \
  and\ \bibinfo {author} {\bibfnamefont {S.~R.}\ \bibnamefont {White}},\
  }\bibfield  {title} {\enquote {\bibinfo {title} {{Spin-liquid ground state of
  the S = 1/2 kagome Heisenberg antiferromagnet.}}}\ }\href@noop {} {\bibfield
  {journal} {\bibinfo  {journal} {Science}\ }\textbf {\bibinfo {volume}
  {332}},\ \bibinfo {pages} {1173} (\bibinfo {year} {2011})}\BibitemShut
  {NoStop}%
\bibitem [{\citenamefont {Depenbrock}\ \emph {et~al.}(2012)\citenamefont
  {Depenbrock}, \citenamefont {McCulloch},\ and\ \citenamefont
  {Schollw\"{o}ck}}]{Depenbrock2012}%
  \BibitemOpen
  \bibfield  {author} {\bibinfo {author} {\bibfnamefont {S.}~\bibnamefont
  {Depenbrock}}, \bibinfo {author} {\bibfnamefont {I.~P.}\ \bibnamefont
  {McCulloch}}, \ and\ \bibinfo {author} {\bibfnamefont {U.}~\bibnamefont
  {Schollw\"{o}ck}},\ }\bibfield  {title} {\enquote {\bibinfo {title} {{Nature
  of the Spin-Liquid Ground State of the S=1/2 Heisenberg Model on the Kagome
  Lattice}},}\ }\href@noop {} {\bibfield  {journal} {\bibinfo  {journal} {Phys.
  Rev. Lett.}\ }\textbf {\bibinfo {volume} {109}},\ \bibinfo {pages} {067201}
  (\bibinfo {year} {2012})}\BibitemShut {NoStop}%
\bibitem [{\citenamefont {{Liao}}\ \emph {et~al.}(2016)\citenamefont {{Liao}},
  \citenamefont {{Xie}}, \citenamefont {{Chen}}, \citenamefont {{Liu}},
  \citenamefont {{Xie}}, \citenamefont {{Huang}}, \citenamefont {{Normand}},\
  and\ \citenamefont {{Xiang}}}]{Liao2016}%
  \BibitemOpen
  \bibfield  {author} {\bibinfo {author} {\bibfnamefont {H.~J.}\ \bibnamefont
  {{Liao}}}, \bibinfo {author} {\bibfnamefont {Z.~Y.}\ \bibnamefont {{Xie}}},
  \bibinfo {author} {\bibfnamefont {J.}~\bibnamefont {{Chen}}}, \bibinfo
  {author} {\bibfnamefont {Z.~Y.}\ \bibnamefont {{Liu}}}, \bibinfo {author}
  {\bibfnamefont {H.~D.}\ \bibnamefont {{Xie}}}, \bibinfo {author}
  {\bibfnamefont {R.~Z.}\ \bibnamefont {{Huang}}}, \bibinfo {author}
  {\bibfnamefont {B.}~\bibnamefont {{Normand}}}, \ and\ \bibinfo {author}
  {\bibfnamefont {T.}~\bibnamefont {{Xiang}}},\ }\bibfield  {title} {\enquote
  {\bibinfo {title} {{Gapless spin-liquid ground state in the $S = 1/2$ kagome
  antiferromagnet}},}\ }\href@noop {} {\bibfield  {journal} {\bibinfo
  {journal} {ArXiv e-prints}\ } (\bibinfo {year} {2016})},\ \Eprint
  {http://arxiv.org/abs/1610.04727} {arXiv:1610.04727 [cond-mat.str-el]}
  \BibitemShut {NoStop}%
\bibitem [{\citenamefont {Nakano}\ and\ \citenamefont
  {Sakai}(2011)}]{Nakano2011}%
  \BibitemOpen
  \bibfield  {author} {\bibinfo {author} {\bibfnamefont {H.}~\bibnamefont
  {Nakano}}\ and\ \bibinfo {author} {\bibfnamefont {T.}~\bibnamefont {Sakai}},\
  }\bibfield  {title} {\enquote {\bibinfo {title} {{Numerical-Diagonalization
  Study of Spin Gap Issue of the Kagome Lattice Heisenberg Antiferromagnet}},}\
  }\href@noop {} {\bibfield  {journal} {\bibinfo  {journal} {J. Phys. Soc.
  Japan}\ }\textbf {\bibinfo {volume} {80}},\ \bibinfo {pages} {053704}
  (\bibinfo {year} {2011})}\BibitemShut {NoStop}%
\bibitem [{\citenamefont {Iqbal}\ \emph {et~al.}(2013)\citenamefont {Iqbal},
  \citenamefont {Becca}, \citenamefont {Sorella},\ and\ \citenamefont
  {Poilblanc}}]{Iqbal2013}%
  \BibitemOpen
  \bibfield  {author} {\bibinfo {author} {\bibfnamefont {Y.}~\bibnamefont
  {Iqbal}}, \bibinfo {author} {\bibfnamefont {F.}~\bibnamefont {Becca}},
  \bibinfo {author} {\bibfnamefont {S.}~\bibnamefont {Sorella}}, \ and\
  \bibinfo {author} {\bibfnamefont {D.}~\bibnamefont {Poilblanc}},\ }\bibfield
  {title} {\enquote {\bibinfo {title} {{Gapless spin-liquid phase in the kagome
  spin-1/2 Heisenberg antiferromagnet}},}\ }\href@noop {} {\bibfield  {journal}
  {\bibinfo  {journal} {Phys. Rev. B}\ }\textbf {\bibinfo {volume} {87}},\
  \bibinfo {pages} {060405(R)} (\bibinfo {year} {2013})}\BibitemShut {NoStop}%
\bibitem [{\citenamefont {Iqbal}\ \emph {et~al.}(2014)\citenamefont {Iqbal},
  \citenamefont {Poilblanc},\ and\ \citenamefont {Becca}}]{Iqbal2014}%
  \BibitemOpen
  \bibfield  {author} {\bibinfo {author} {\bibfnamefont {Y.}~\bibnamefont
  {Iqbal}}, \bibinfo {author} {\bibfnamefont {D.}~\bibnamefont {Poilblanc}}, \
  and\ \bibinfo {author} {\bibfnamefont {F.}~\bibnamefont {Becca}},\ }\bibfield
   {title} {\enquote {\bibinfo {title} {{Vanishing spin gap in a competing
  spin-liquid phase in the kagome Heisenberg antiferromagnet}},}\ }\href@noop
  {} {\bibfield  {journal} {\bibinfo  {journal} {Phys. Rev. B}\ }\textbf
  {\bibinfo {volume} {89}},\ \bibinfo {pages} {020407(R)} (\bibinfo {year}
  {2014})}\BibitemShut {NoStop}%
\bibitem [{\citenamefont {Shores}\ \emph {et~al.}(2005)\citenamefont {Shores},
  \citenamefont {Nytko}, \citenamefont {Bartlett},\ and\ \citenamefont
  {Nocera}}]{Shores2005}%
  \BibitemOpen
  \bibfield  {author} {\bibinfo {author} {\bibfnamefont {M.~P.}\ \bibnamefont
  {Shores}}, \bibinfo {author} {\bibfnamefont {E.~A.}\ \bibnamefont {Nytko}},
  \bibinfo {author} {\bibfnamefont {B.~M.}\ \bibnamefont {Bartlett}}, \ and\
  \bibinfo {author} {\bibfnamefont {D.~G.}\ \bibnamefont {Nocera}},\ }\bibfield
   {title} {\enquote {\bibinfo {title} {{A structurally perfect S = (1/2)
  kagome antiferromagnet.}}}\ }\href@noop {} {\bibfield  {journal} {\bibinfo
  {journal} {J. Am. Chem. Soc.}\ }\textbf {\bibinfo {volume} {127}},\ \bibinfo
  {pages} {13462--13463} (\bibinfo {year} {2005})}\BibitemShut {NoStop}%
\bibitem [{\citenamefont {Mendels}\ and\ \citenamefont
  {Bert}(2011)}]{Mendels2011}%
  \BibitemOpen
  \bibfield  {author} {\bibinfo {author} {\bibfnamefont {P.}~\bibnamefont
  {Mendels}}\ and\ \bibinfo {author} {\bibfnamefont {F.}~\bibnamefont {Bert}},\
  }\bibfield  {title} {\enquote {\bibinfo {title} {{Quantum kagome
  antiferromagnet : ZnCu$_3$(OH)$_6$Cl$_2$}},}\ }\href@noop {} {\bibfield
  {journal} {\bibinfo  {journal} {J. Phys. Conf. Ser.}\ }\textbf {\bibinfo
  {volume} {320}},\ \bibinfo {pages} {012004} (\bibinfo {year}
  {2011})}\BibitemShut {NoStop}%
\bibitem [{\citenamefont {Norman}(2016)}]{Norman2016}%
  \BibitemOpen
  \bibfield  {author} {\bibinfo {author} {\bibfnamefont {M.~R.}\ \bibnamefont
  {Norman}},\ }\bibfield  {title} {\enquote {\bibinfo {title}
  {\textit{Colloquium} : Herbertsmithite and the search for the quantum spin
  liquid},}\ }\href {\doibase 10.1103/RevModPhys.88.041002} {\bibfield
  {journal} {\bibinfo  {journal} {Rev. Mod. Phys.}\ }\textbf {\bibinfo {volume}
  {88}},\ \bibinfo {pages} {041002} (\bibinfo {year} {2016})}\BibitemShut
  {NoStop}%
\bibitem [{\citenamefont {Olariu}\ \emph {et~al.}(2008)\citenamefont {Olariu},
  \citenamefont {Mendels}, \citenamefont {Bert}, \citenamefont {Duc},
  \citenamefont {Trombe}, \citenamefont {de~Vries},\ and\ \citenamefont
  {Harrison}}]{Olariu2008}%
  \BibitemOpen
  \bibfield  {author} {\bibinfo {author} {\bibfnamefont {A.}~\bibnamefont
  {Olariu}}, \bibinfo {author} {\bibfnamefont {P.}~\bibnamefont {Mendels}},
  \bibinfo {author} {\bibfnamefont {F.}~\bibnamefont {Bert}}, \bibinfo {author}
  {\bibfnamefont {F.}~\bibnamefont {Duc}}, \bibinfo {author} {\bibfnamefont
  {J.~C.}\ \bibnamefont {Trombe}}, \bibinfo {author} {\bibfnamefont {M.~A.}\
  \bibnamefont {de~Vries}}, \ and\ \bibinfo {author} {\bibfnamefont
  {A.}~\bibnamefont {Harrison}},\ }\bibfield  {title} {\enquote {\bibinfo
  {title} {{$^{17}$O NMR Study of the Intrinsic Magnetic Susceptibility and
  Spin Dynamics of the Quantum Kagome Antiferromagnet
  ZnCu$_3$(OH)$_6$Cl$_2$}},}\ }\href@noop {} {\bibfield  {journal} {\bibinfo
  {journal} {Phys. Rev. Lett.}\ }\textbf {\bibinfo {volume} {100}},\ \bibinfo
  {pages} {087202} (\bibinfo {year} {2008})}\BibitemShut {NoStop}%
\bibitem [{\citenamefont {Fu}\ \emph {et~al.}(2015)\citenamefont {Fu},
  \citenamefont {Imai}, \citenamefont {Han},\ and\ \citenamefont
  {Lee}}]{Fu2015}%
  \BibitemOpen
  \bibfield  {author} {\bibinfo {author} {\bibfnamefont {M.}~\bibnamefont
  {Fu}}, \bibinfo {author} {\bibfnamefont {T.}~\bibnamefont {Imai}}, \bibinfo
  {author} {\bibfnamefont {T.~H.}\ \bibnamefont {Han}}, \ and\ \bibinfo
  {author} {\bibfnamefont {Y.~S.}\ \bibnamefont {Lee}},\ }\bibfield  {title}
  {\enquote {\bibinfo {title} {{Evidence for a gapped spin-liquid ground state
  in a kagome Heisenberg antiferromagnet}},}\ }\href@noop {} {\bibfield
  {journal} {\bibinfo  {journal} {Science}\ }\textbf {\bibinfo {volume}
  {350}},\ \bibinfo {pages} {655} (\bibinfo {year} {2015})}\BibitemShut
  {NoStop}%
\bibitem [{\citenamefont {Han}\ \emph {et~al.}(2012)\citenamefont {Han},
  \citenamefont {Helton}, \citenamefont {Chu}, \citenamefont {Nocera},
  \citenamefont {Rodriguez-Rivera}, \citenamefont {Broholm},\ and\
  \citenamefont {Lee}}]{Han2012a}%
  \BibitemOpen
  \bibfield  {author} {\bibinfo {author} {\bibfnamefont {T.~H.}\ \bibnamefont
  {Han}}, \bibinfo {author} {\bibfnamefont {J.~S.}\ \bibnamefont {Helton}},
  \bibinfo {author} {\bibfnamefont {S.}~\bibnamefont {Chu}}, \bibinfo {author}
  {\bibfnamefont {D.~G.}\ \bibnamefont {Nocera}}, \bibinfo {author}
  {\bibfnamefont {J.~A.}\ \bibnamefont {Rodriguez-Rivera}}, \bibinfo {author}
  {\bibfnamefont {C.}~\bibnamefont {Broholm}}, \ and\ \bibinfo {author}
  {\bibfnamefont {Y.~S.}\ \bibnamefont {Lee}},\ }\bibfield  {title} {\enquote
  {\bibinfo {title} {{Fractionalized excitations in the spin-liquid state of a
  kagome-lattice antiferromagnet}},}\ }\href@noop {} {\bibfield  {journal}
  {\bibinfo  {journal} {Nature}\ }\textbf {\bibinfo {volume} {492}},\ \bibinfo
  {pages} {406--410} (\bibinfo {year} {2012})}\BibitemShut {NoStop}%
\bibitem [{\citenamefont {Schaffer}\ \emph {et~al.}(2017)\citenamefont
  {Schaffer}, \citenamefont {Huh}, \citenamefont {Hwang},\ and\ \citenamefont
  {Kim}}]{Schaffer17}%
  \BibitemOpen
  \bibfield  {author} {\bibinfo {author} {\bibfnamefont {Robert}\ \bibnamefont
  {Schaffer}}, \bibinfo {author} {\bibfnamefont {Yejin}\ \bibnamefont {Huh}},
  \bibinfo {author} {\bibfnamefont {Kyusung}\ \bibnamefont {Hwang}}, \ and\
  \bibinfo {author} {\bibfnamefont {Yong~Baek}\ \bibnamefont {Kim}},\
  }\bibfield  {title} {\enquote {\bibinfo {title} {Quantum spin liquid in a
  breathing kagome lattice},}\ }\href {\doibase 10.1103/PhysRevB.95.054410}
  {\bibfield  {journal} {\bibinfo  {journal} {Phys. Rev. B}\ }\textbf {\bibinfo
  {volume} {95}},\ \bibinfo {pages} {054410} (\bibinfo {year}
  {2017})}\BibitemShut {NoStop}%
\bibitem [{\citenamefont {Okamoto}\ \emph {et~al.}(2013)\citenamefont
  {Okamoto}, \citenamefont {Nilsen}, \citenamefont {Attfield},\ and\
  \citenamefont {Hiroi}}]{Okamoto2013}%
  \BibitemOpen
  \bibfield  {author} {\bibinfo {author} {\bibfnamefont {Y.}~\bibnamefont
  {Okamoto}}, \bibinfo {author} {\bibfnamefont {G.~J.}\ \bibnamefont {Nilsen}},
  \bibinfo {author} {\bibfnamefont {J.~P.}\ \bibnamefont {Attfield}}, \ and\
  \bibinfo {author} {\bibfnamefont {Z.}~\bibnamefont {Hiroi}},\ }\bibfield
  {title} {\enquote {\bibinfo {title} {{Breathing Pyrochlore Lattice Realized
  in $A$-Site Ordered Spinel Oxides LiGaCr$_4$O$_8$ and LiInCr$_4$O$_8$}},}\
  }\href {\doibase 10.1103/PhysRevLett.110.097203} {\bibfield  {journal}
  {\bibinfo  {journal} {Phys. Rev. Lett.}\ }\textbf {\bibinfo {volume} {110}},\
  \bibinfo {pages} {097203} (\bibinfo {year} {2013})}\BibitemShut {NoStop}%
\bibitem [{\citenamefont {Aidoudi}\ \emph {et~al.}(2011)\citenamefont
  {Aidoudi}, \citenamefont {Aldous}, \citenamefont {Goff}, \citenamefont
  {Slawin}, \citenamefont {Attfield}, \citenamefont {Morris},\ and\
  \citenamefont {Lightfoot}}]{Aidoudi2011}%
  \BibitemOpen
  \bibfield  {author} {\bibinfo {author} {\bibfnamefont {F.~H.}\ \bibnamefont
  {Aidoudi}}, \bibinfo {author} {\bibfnamefont {D.~W.}\ \bibnamefont {Aldous}},
  \bibinfo {author} {\bibfnamefont {R.~J.}\ \bibnamefont {Goff}}, \bibinfo
  {author} {\bibfnamefont {A.~M.~Z.}\ \bibnamefont {Slawin}}, \bibinfo {author}
  {\bibfnamefont {J.~P.}\ \bibnamefont {Attfield}}, \bibinfo {author}
  {\bibfnamefont {R.~E.}\ \bibnamefont {Morris}}, \ and\ \bibinfo {author}
  {\bibfnamefont {P.}~\bibnamefont {Lightfoot}},\ }\bibfield  {title} {\enquote
  {\bibinfo {title} {{An ionothermally prepared S = 1/2 vanadium oxyfluoride
  kagome lattice.}}}\ }\href@noop {} {\bibfield  {journal} {\bibinfo  {journal}
  {Nat. Chem.}\ }\textbf {\bibinfo {volume} {3}},\ \bibinfo {pages} {801--806}
  (\bibinfo {year} {2011})}\BibitemShut {NoStop}%
\bibitem [{Note1()}]{Note1}%
  \BibitemOpen
  \bibinfo {note} {The VOF$_5$ octahedron is strongly distorted with a distance
  $d_{V^{4+}-O}=1.58$~\r A{} much shorter than the distances $d_{V^{4+}-F}$ all
  in the range $1.95-2.15$~\r A{} which lifts the $t_{2g}$ orbitals degeneracy
  and favors the $d_{xy}$ one~\cite {Aidoudi2011}.}\BibitemShut {Stop}%
\bibitem [{\citenamefont {Clark}\ \emph {et~al.}(2013)\citenamefont {Clark},
  \citenamefont {Orain}, \citenamefont {Bert}, \citenamefont {{De Vries}},
  \citenamefont {Aidoudi}, \citenamefont {Morris}, \citenamefont {Lightfoot},
  \citenamefont {Lord}, \citenamefont {Telling}, \citenamefont {Bonville},
  \citenamefont {Attfield}, \citenamefont {Mendels},\ and\ \citenamefont
  {Harrison}}]{Clark2013}%
  \BibitemOpen
  \bibfield  {author} {\bibinfo {author} {\bibfnamefont {L.}~\bibnamefont
  {Clark}}, \bibinfo {author} {\bibfnamefont {J.~C.}\ \bibnamefont {Orain}},
  \bibinfo {author} {\bibfnamefont {F.}~\bibnamefont {Bert}}, \bibinfo {author}
  {\bibfnamefont {M.~A.}\ \bibnamefont {{De Vries}}}, \bibinfo {author}
  {\bibfnamefont {F.~H.}\ \bibnamefont {Aidoudi}}, \bibinfo {author}
  {\bibfnamefont {R.~E.}\ \bibnamefont {Morris}}, \bibinfo {author}
  {\bibfnamefont {P.}~\bibnamefont {Lightfoot}}, \bibinfo {author}
  {\bibfnamefont {J.~S.}\ \bibnamefont {Lord}}, \bibinfo {author}
  {\bibfnamefont {M.~T.~F.}\ \bibnamefont {Telling}}, \bibinfo {author}
  {\bibfnamefont {P.}~\bibnamefont {Bonville}}, \bibinfo {author}
  {\bibfnamefont {J.~P.}\ \bibnamefont {Attfield}}, \bibinfo {author}
  {\bibfnamefont {P.}~\bibnamefont {Mendels}}, \ and\ \bibinfo {author}
  {\bibfnamefont {A.}~\bibnamefont {Harrison}},\ }\bibfield  {title} {\enquote
  {\bibinfo {title} {{Gapless Spin Liquid Ground State in the S = 1/2 Vanadium
  Oxyfluoride Kagome Antiferromagnet
  [NH$_4$]$_2$[C$_7$H$_{14}$N][V$_7$O$_6$F$_{18}$]}},}\ }\href@noop {}
  {\bibfield  {journal} {\bibinfo  {journal} {Phys. Rev. Lett.}\ }\textbf
  {\bibinfo {volume} {110}},\ \bibinfo {pages} {207208} (\bibinfo {year}
  {2013})}\BibitemShut {NoStop}%
\bibitem [{\citenamefont {Orain}\ \emph {et~al.}(2014)\citenamefont {Orain},
  \citenamefont {Clark}, \citenamefont {Bert}, \citenamefont {Mendels},
  \citenamefont {Attfield}, \citenamefont {Aidoudi}, \citenamefont {Morris},
  \citenamefont {Lightfoot}, \citenamefont {Amato},\ and\ \citenamefont
  {Baines}}]{Orain2014}%
  \BibitemOpen
  \bibfield  {author} {\bibinfo {author} {\bibfnamefont {J.~C.}\ \bibnamefont
  {Orain}}, \bibinfo {author} {\bibfnamefont {L.}~\bibnamefont {Clark}},
  \bibinfo {author} {\bibfnamefont {F.}~\bibnamefont {Bert}}, \bibinfo {author}
  {\bibfnamefont {P.}~\bibnamefont {Mendels}}, \bibinfo {author} {\bibfnamefont
  {J.~P.}\ \bibnamefont {Attfield}}, \bibinfo {author} {\bibfnamefont {F.~H.}\
  \bibnamefont {Aidoudi}}, \bibinfo {author} {\bibfnamefont {R.~E.}\
  \bibnamefont {Morris}}, \bibinfo {author} {\bibfnamefont {P.}~\bibnamefont
  {Lightfoot}}, \bibinfo {author} {\bibfnamefont {A.}~\bibnamefont {Amato}}, \
  and\ \bibinfo {author} {\bibfnamefont {C.}~\bibnamefont {Baines}},\
  }\bibfield  {title} {\enquote {\bibinfo {title} {{MuSR study of a quantum
  spin liquid candidate : the S=1/2 vanadium oxyfluoride kagome
  antiferromagnet}},}\ }\href@noop {} {\bibfield  {journal} {\bibinfo
  {journal} {J. Phys. Conf. Ser.}\ }\textbf {\bibinfo {volume} {551}},\
  \bibinfo {pages} {012004} (\bibinfo {year} {2014})}\BibitemShut {NoStop}%
\bibitem [{\citenamefont {Griffin}\ \emph {et~al.}(2012)\citenamefont
  {Griffin}, \citenamefont {Clark}, \citenamefont {Seymour}, \citenamefont
  {Aldous}, \citenamefont {Dawson}, \citenamefont {Iuga}, \citenamefont
  {Morris},\ and\ \citenamefont {Ashbrook}}]{Griffin2012}%
  \BibitemOpen
  \bibfield  {author} {\bibinfo {author} {\bibfnamefont {J.M.}\ \bibnamefont
  {Griffin}}, \bibinfo {author} {\bibfnamefont {L.}~\bibnamefont {Clark}},
  \bibinfo {author} {\bibfnamefont {V.R.}\ \bibnamefont {Seymour}}, \bibinfo
  {author} {\bibfnamefont {D.W.}\ \bibnamefont {Aldous}}, \bibinfo {author}
  {\bibfnamefont {D.M.}\ \bibnamefont {Dawson}}, \bibinfo {author}
  {\bibfnamefont {D.}~\bibnamefont {Iuga}}, \bibinfo {author} {\bibfnamefont
  {R.E.}\ \bibnamefont {Morris}}, \ and\ \bibinfo {author} {\bibfnamefont
  {S.E.}\ \bibnamefont {Ashbrook}},\ }\bibfield  {title} {\enquote {\bibinfo
  {title} {{Ionothermal $^{17}$O enrichment of oxides using microlitre
  quantities of labelled water}},}\ }\href {\doibase 10.1039/C2SC20155K}
  {\bibfield  {journal} {\bibinfo  {journal} {Chem. Sci.}\ }\textbf {\bibinfo
  {volume} {3}},\ \bibinfo {pages} {2293--2300} (\bibinfo {year}
  {2012})}\BibitemShut {NoStop}%
\bibitem [{Note2()}]{Note2}%
  \BibitemOpen
  \bibinfo {note} {From a powder spectrum simulation, we extract an average
  quadrupolar frequency $\nu _Q=370$~kHz with a small distribution $\Delta \nu
  _Q=10$~kHz and a maximal anisotropy of the quadrupolar tensor $\eta <0.05$.
  The shift tensor $\protect \mathaccentV {bar}016{\protect \mathbf {K}}$ is
  almost isotropic with $K_{x,y}/K_z=0.87$, negligible in view of the low $T$
  broadening.}\BibitemShut {Stop}%
\bibitem [{SI()}]{SI}%
  \BibitemOpen
  \href@noop {} {}\bibinfo {note} {See Supplemental Material for details on the
  determination of the NMR parameters, on the series expansion method, fits of
  the susceptibility and linewidth of the NMR spectra.}\BibitemShut {Stop}%
\bibitem [{\citenamefont {Bernu}\ and\ \citenamefont
  {Lhuillier}(2015)}]{Bernu2015}%
  \BibitemOpen
  \bibfield  {author} {\bibinfo {author} {\bibfnamefont {B.}~\bibnamefont
  {Bernu}}\ and\ \bibinfo {author} {\bibfnamefont {C.}~\bibnamefont
  {Lhuillier}},\ }\bibfield  {title} {\enquote {\bibinfo {title} {{Spin
  Susceptibility of Quantum Magnets from High to Low Temperatures}},}\
  }\href@noop {} {\bibfield  {journal} {\bibinfo  {journal} {Phys. Rev. Lett.}\
  }\textbf {\bibinfo {volume} {114}},\ \bibinfo {pages} {057201} (\bibinfo
  {year} {2015})}\BibitemShut {NoStop}%
\bibitem [{\citenamefont {Nishimoto}\ \emph {et~al.}(2013)\citenamefont
  {Nishimoto}, \citenamefont {Shibata},\ and\ \citenamefont
  {Hotta}}]{Nishimoto13}%
  \BibitemOpen
  \bibfield  {author} {\bibinfo {author} {\bibfnamefont {Satoshi}\ \bibnamefont
  {Nishimoto}}, \bibinfo {author} {\bibfnamefont {Naokazu}\ \bibnamefont
  {Shibata}}, \ and\ \bibinfo {author} {\bibfnamefont {Chisa}\ \bibnamefont
  {Hotta}},\ }\bibfield  {title} {\enquote {\bibinfo {title} {Controlling
  frustrated liquids and solids with an applied field in a kagome heisenberg
  antiferromagnet},}\ }\href {http://dx.doi.org/10.1038/ncomms3287} {\bibfield
  {journal} {\bibinfo  {journal} {Nature Communications}\ }\textbf {\bibinfo
  {volume} {4}},\ \bibinfo {pages} {2287--} (\bibinfo {year}
  {2013})}\BibitemShut {NoStop}%
\bibitem [{\citenamefont {{He}}\ \emph {et~al.}(2016)\citenamefont {{He}},
  \citenamefont {{Zaletel}}, \citenamefont {{Oshikawa}},\ and\ \citenamefont
  {{Pollmann}}}]{He16}%
  \BibitemOpen
  \bibfield  {author} {\bibinfo {author} {\bibfnamefont {Y.-C.}\ \bibnamefont
  {{He}}}, \bibinfo {author} {\bibfnamefont {M.~P.}\ \bibnamefont {{Zaletel}}},
  \bibinfo {author} {\bibfnamefont {M.}~\bibnamefont {{Oshikawa}}}, \ and\
  \bibinfo {author} {\bibfnamefont {F.}~\bibnamefont {{Pollmann}}},\ }\bibfield
   {title} {\enquote {\bibinfo {title} {{Signatures of Dirac cones in a DMRG
  study of the Kagome Heisenberg model}},}\ }\href@noop {} {\bibfield
  {journal} {\bibinfo  {journal} {ArXiv e-prints}\ } (\bibinfo {year}
  {2016})},\ \Eprint {http://arxiv.org/abs/1611.06238} {arXiv:1611.06238
  [cond-mat.str-el]} \BibitemShut {NoStop}%
\bibitem [{\citenamefont {Suter}\ \emph {et~al.}(1998)\citenamefont {Suter},
  \citenamefont {Mali}, \citenamefont {Roos},\ and\ \citenamefont
  {Brinkmann}}]{Suter1998}%
  \BibitemOpen
  \bibfield  {author} {\bibinfo {author} {\bibfnamefont {A.}~\bibnamefont
  {Suter}}, \bibinfo {author} {\bibfnamefont {M.}~\bibnamefont {Mali}},
  \bibinfo {author} {\bibfnamefont {J.}~\bibnamefont {Roos}}, \ and\ \bibinfo
  {author} {\bibfnamefont {D.}~\bibnamefont {Brinkmann}},\ }\bibfield  {title}
  {\enquote {\bibinfo {title} {{Mixed magnetic and quadrupolar relaxation in
  the presence of a dominant static Zeeman Hamiltonian}},}\ }\href {\doibase
  10.1088/0953-8984/10/26/022} {\bibfield  {journal} {\bibinfo  {journal} {J.
  Phys. Condens. Matter}\ }\textbf {\bibinfo {volume} {10}},\ \bibinfo {pages}
  {5977--5994} (\bibinfo {year} {1998})}\BibitemShut {NoStop}%
\bibitem [{\citenamefont {Mila}(1998)}]{Mila1998}%
  \BibitemOpen
  \bibfield  {author} {\bibinfo {author} {\bibfnamefont {F.}~\bibnamefont
  {Mila}},\ }\bibfield  {title} {\enquote {\bibinfo {title} {{Low-Energy Sector
  of the S=1/2 Kagome Antiferromagnet}},}\ }\href@noop {} {\bibfield  {journal}
  {\bibinfo  {journal} {Phys. Rev. Lett.}\ }\textbf {\bibinfo {volume} {81}},\
  \bibinfo {pages} {2356} (\bibinfo {year} {1998})}\BibitemShut {NoStop}%
\bibitem [{\citenamefont {Mambrini}\ and\ \citenamefont
  {Mila}(2000)}]{Mambrini2000}%
  \BibitemOpen
  \bibfield  {author} {\bibinfo {author} {\bibfnamefont {M.}~\bibnamefont
  {Mambrini}}\ and\ \bibinfo {author} {\bibfnamefont {F.}~\bibnamefont
  {Mila}},\ }\bibfield  {title} {\enquote {\bibinfo {title} {{RVB description
  of the low-energy singlets of the spin 1/2 kagome antiferromagnet}},}\
  }\href@noop {} {\bibfield  {journal} {\bibinfo  {journal} {Eur. Phys. J. B}\
  }\textbf {\bibinfo {volume} {17}},\ \bibinfo {pages} {651--659} (\bibinfo
  {year} {2000})}\BibitemShut {NoStop}%
\bibitem [{\citenamefont {Zhitomirsky}(2005)}]{Zhitomirsky2005a}%
  \BibitemOpen
  \bibfield  {author} {\bibinfo {author} {\bibfnamefont {M.E.}~\bibnamefont
  {Zhitomirsky}},\ }\bibfield  {title} {\enquote {\bibinfo {title} {{Effective
  quantum dimer model for trimerized kagome antiferromagnet}},}\ }\href@noop {}
  {\bibfield  {journal} {\bibinfo  {journal} {Phys. Rev. B}\ }\textbf {\bibinfo
  {volume} {71}},\ \bibinfo {pages} {214413} (\bibinfo {year}
  {2005})}\BibitemShut {NoStop}%
\bibitem [{\citenamefont {Kolezhuk}\ \emph {et~al.}(2006)\citenamefont
  {Kolezhuk}, \citenamefont {Sachdev}, \citenamefont {Biswas},\ and\
  \citenamefont {Chen}}]{Kolezhuk2006}%
  \BibitemOpen
  \bibfield  {author} {\bibinfo {author} {\bibfnamefont {A.}~\bibnamefont
  {Kolezhuk}}, \bibinfo {author} {\bibfnamefont {S.}~\bibnamefont {Sachdev}},
  \bibinfo {author} {\bibfnamefont {R.R.}\ \bibnamefont {Biswas}}, \ and\
  \bibinfo {author} {\bibfnamefont {P.}~\bibnamefont {Chen}},\ }\bibfield
  {title} {\enquote {\bibinfo {title} {Theory of quantum impurities in spin
  liquids},}\ }\href {\doibase 10.1103/PhysRevB.74.165114} {\bibfield
  {journal} {\bibinfo  {journal} {Phys. Rev. B}\ }\textbf {\bibinfo {volume}
  {74}},\ \bibinfo {pages} {165114} (\bibinfo {year} {2006})}\BibitemShut
  {NoStop}%
\bibitem [{\citenamefont {Gomilsek}\ \emph {et~al.}(2016)\citenamefont
  {Gomilsek}, \citenamefont {Klanjsek}, \citenamefont {Pregelj}, \citenamefont
  {Luetkens}, \citenamefont {Li}, \citenamefont {Zhang},\ and\ \citenamefont
  {Zorko}}]{Gomilsek2016}%
  \BibitemOpen
  \bibfield  {author} {\bibinfo {author} {\bibfnamefont {M.}~\bibnamefont
  {Gomilsek}}, \bibinfo {author} {\bibfnamefont {M.}~\bibnamefont {Klanjsek}},
  \bibinfo {author} {\bibfnamefont {M.}~\bibnamefont {Pregelj}}, \bibinfo
  {author} {\bibfnamefont {H.}~\bibnamefont {Luetkens}}, \bibinfo {author}
  {\bibfnamefont {Y.}~\bibnamefont {Li}}, \bibinfo {author} {\bibfnamefont
  {Q.~M.}\ \bibnamefont {Zhang}}, \ and\ \bibinfo {author} {\bibfnamefont
  {A.}~\bibnamefont {Zorko}},\ }\bibfield  {title} {\enquote {\bibinfo {title}
  {$\ensuremath{\mu}\mathrm{SR}$ insight into the impurity problem in quantum
  kagome antiferromagnets},}\ }\href {\doibase 10.1103/PhysRevB.94.024438}
  {\bibfield  {journal} {\bibinfo  {journal} {Phys. Rev. B}\ }\textbf {\bibinfo
  {volume} {94}},\ \bibinfo {pages} {024438} (\bibinfo {year}
  {2016})}\BibitemShut {NoStop}%
\bibitem [{\citenamefont {Han}\ \emph {et~al.}(2016)\citenamefont {Han},
  \citenamefont {Norman}, \citenamefont {Wen}, \citenamefont
  {Rodriguez-Rivera}, \citenamefont {Helton}, \citenamefont {Broholm},\ and\
  \citenamefont {Lee}}]{Han2016}%
  \BibitemOpen
  \bibfield  {author} {\bibinfo {author} {\bibfnamefont {T.-H.}\ \bibnamefont
  {Han}}, \bibinfo {author} {\bibfnamefont {M.R.}\ \bibnamefont {Norman}},
  \bibinfo {author} {\bibfnamefont {J.-J.}\ \bibnamefont {Wen}}, \bibinfo
  {author} {\bibfnamefont {J.A.}\ \bibnamefont {Rodriguez-Rivera}}, \bibinfo
  {author} {\bibfnamefont {J.S.}\ \bibnamefont {Helton}}, \bibinfo {author}
  {\bibfnamefont {C.}~\bibnamefont {Broholm}}, \ and\ \bibinfo {author}
  {\bibfnamefont {Y.S.}\ \bibnamefont {Lee}},\ }\bibfield  {title} {\enquote
  {\bibinfo {title} {Correlated impurities and intrinsic spin-liquid physics in
  the kagome material herbertsmithite},}\ }\href {\doibase
  10.1103/PhysRevB.94.060409} {\bibfield  {journal} {\bibinfo  {journal} {Phys.
  Rev. B}\ }\textbf {\bibinfo {volume} {94}},\ \bibinfo {pages} {060409}
  (\bibinfo {year} {2016})}\BibitemShut {NoStop}%
\bibitem [{\citenamefont {Clark}\ \emph {et~al.}(2015)\citenamefont {Clark},
  \citenamefont {Aidoudi}, \citenamefont {Black}, \citenamefont {Arachchige},
  \citenamefont {Slawin}, \citenamefont {Morris},\ and\ \citenamefont
  {Lightfoot}}]{Clark2015}%
  \BibitemOpen
  \bibfield  {author} {\bibinfo {author} {\bibfnamefont {L.}~\bibnamefont
  {Clark}}, \bibinfo {author} {\bibfnamefont {F.~H.}\ \bibnamefont {Aidoudi}},
  \bibinfo {author} {\bibfnamefont {C.}~\bibnamefont {Black}}, \bibinfo
  {author} {\bibfnamefont {K.~S.~A.}\ \bibnamefont {Arachchige}}, \bibinfo
  {author} {\bibfnamefont {A.~M.~Z.}\ \bibnamefont {Slawin}}, \bibinfo {author}
  {\bibfnamefont {R.~E.}\ \bibnamefont {Morris}}, \ and\ \bibinfo {author}
  {\bibfnamefont {P.}~\bibnamefont {Lightfoot}},\ }\bibfield  {title} {\enquote
  {\bibinfo {title} {{Extending the Familly of V$^{4+}$ S=1/2 Kagome
  Antiferromagnets}},}\ }\href@noop {} {\bibfield  {journal} {\bibinfo
  {journal} {Angew. Chem. Int. Ed.}\ }\textbf {\bibinfo {volume} {54}}
  (\bibinfo {year} {2015})}\BibitemShut {NoStop}%
\end{thebibliography}
%

\end{document}